\documentclass[useAMS,usenatbib]{mn2e}
\usepackage{times}
\usepackage{graphicx}
\usepackage{amssymb}

\title[Mergers]{A Surprisingly High Pair Fraction for Extremely Massive Galaxies at z $\sim$ 3 in the GOODS NICMOS Survey}
\author[Asa F. L. Bluck et al.]{Asa~F.~L.~Bluck$^1$, Christopher~J.~Conselice$^1$, Rychard~J.~Bouwens$^2$, Emanuele~Daddi$^3$, \newauthor Mark~Dickinson$^4$, Casey~Papovich$^5$, Haojing~Yan$^6$ \footnotemark[0]\\
\\$^1$ University of Nottingham, School of Physics and Astronomy, Nottingham NG7 2RD, UK
\\$^2$ Astronomy Department, University of California, Santa Cruz, CA 95064, USA
\\$^3$ Laboratoire AIM, CEA/DSM-CNRS-Universit´e Paris Diderot, Irfu/SAp, Orme des Merisiers, F-91191, Gif-sur-Yvette, France
\\$^4$ National Optical Astronomy Observatory, Tucson, AZ 85719, USA
\\$^5$ Astronomy Department, Texas A \& M University, TX, USA
\\$^6$ The Observatories of the Carnegie Institution of Washington, Pasedena, CA 91101, USA}

\begin{document}

\maketitle

\begin{abstract}
We calculate the major pair fraction and derive the major merger fraction and rate for 82 massive ($M_{*}>10^{11}M_{\odot}$) galaxies at $1.7 < z < 3.0$ utilising deep HST NICMOS data taken in the GOODS North and South fields. For the first time, our NICMOS data provides imaging with sufficient angular resolution and depth to collate a sufficiently large sample of massive galaxies at z $>$ 1.5 to reliably measure their pair fraction history. We find strong evidence that the pair fraction of massive galaxies evolves with redshift. We calculate a pair fraction of $f_{m}$ = 0.29 +/- 0.06 for our whole sample at $1.7 < z < 3.0$. Specifically, we fit a power law function of the form $f_{m}=f_{0}(1+z)^{m}$ to a combined sample of low redshift data from Conselice et al. (2007) and recently acquired high redshift data from the GOODS NICMOS Survey. We find a best fit to the free parameters of $f_{0}$ = 0.008 +/- 0.003 and $m$ = 3.0 +/- 0.4. We go on to fit a theoretically motivated Press-Schechter curve to this data. This Press-Schechter fit, and the data, show no sign of levelling off or turning over, implying that the merger fraction of massive galaxies continues to rise with redshift out to z $\sim$ 3. Since previous work has established that the merger fraction for lower mass galaxies turns over at z $\sim$ 1.5 - 2.0, this is evidence that higher mass galaxies experience more mergers earlier than their lower mass counterparts, i.e. a galaxy assembly downsizing. Finally, we calculate a merger rate at z = 2.6 of $\Re$ $<$ 5 $\times$ 10$^{5}$ Gpc$^{-3}$ Gyr$^{-1}$, which experiences no significant change to $\Re$ $<$ 1.2 $\times$ 10$^{5}$ Gpc$^{-3}$ Gyr$^{-1}$ at z = 0.5. This corresponds to an average $M_{*}>10^{11}M_{\odot}$ galaxy experiencing 1.7 +/- 0.5 mergers between z = 3 and z = 0.

\end{abstract}


\section{Introduction}
Hierarchical assembly is the established leading theory for the evolution of structure in the universe from the viewpoint of Cold Dark Matter (CDM) theory. This states that larger structures form from the merging of smaller structures. Thus CDM halos grow in size, with the largest structures forming latest in the history of the universe. Galaxies are often thought to form principally in line with halo mergers. If they do, it is likely that merging of smaller galaxies to create more massive ones is the major factor in the evolution of galaxies over cosmic time. Alternatives to a merger history of massive galaxies include rapid collapse mechanisms, whereby galaxies form over very short time-scales in the early universe, and then do not significantly merge during the rest of their lifetime.

The aim of this paper is to address the question: what mechanism drives galaxy evolution with regards to the most massive ($M_{*}>10^{11}M_{\odot}$) galaxies in the universe? We can observationally test this by calculating the merger fraction of massive galaxies at different redshifts. Previous work by Conselice et al. (2007) explores this problem for similar mass galaxies out to z $\sim$ 1.4, using morphological techniques. We extend this work using a close-pair method to z $\sim$ 3, using data from the GOODS NICMOS Survey.

There are several different ways to locate merging galaxies which roughly relate to the stage of the merger. The most direct method is to look at morphological disturbances in galaxies, e.g. Conselice et al. (2003, 2006 and 2008). In this approach one selects galaxies with asymmetries, or distortions in their morphologies, above a certain threshold, and define these as merging systems. This can be carried out by eye, or via computational methods, e.g. CAS, Gini and $M_{20}$ parameters (see Conselice 2003, 2006, Conselice at al. 2008 and Lotz et al. 2008a). Alternatively, one could also look in principle for secondary features such as an abrupt increase in star formation rate. This is because the merging of gas rich galaxies will often give rise to an increase in star formation, and hence a distinctive spectral signature, see e.g. Lin et al. (2008) and Mathis et al. (2005).

Also one can look for galaxies in close proximity and count the number of galaxies in apparent pairs in a given sample. This pair counting method requires much less angular resolution than the morphological approaches, and can be extended to higher redshift more directly than the others. The close pair method does not precisely trace merging, however, but rather looks for likely potential future mergers. Thus, to go from a close pair fraction to a merger fraction requires that assumptions be made. Ultimately, the pair count method measures mergers before they happen, the morphological approach measures merging while it happens, and methods involving increased star formation rates trace merging during and after the event.

Merging is already known to be of paramount importance in driving the rate of star formation, the formation and evolution of supermassive black holes, and as a mechanism for increasing galaxy mass. Further, mergers can be used to trace the formation history of galaxies, probing where, when, and how galaxies form. Conselice et al. (2007) and Rawat et al. (2008) find that massive galaxies ($M_{*}>10^{11}M_{\odot}$) out to z $\sim$ 1.5 have an increasing merger fraction with redshift. It has been noted for lower mass systems that there is a levelling off and eventual turn around in the merger fraction at $z < 2$ (see Conselice et al. 2008). In order to probe whether the same levelling off and turn around in the merger fraction occurs for $M_{*}>10^{11}M_{\odot}$ galaxies, it is necessary to explore the merger fraction of these massive galaxies at higher redshift.

Consequently, we have collated a sample of 82 galaxies with $M_{*}>10^{11}M_{\odot}$, selected from the GOODS North and GOODS South fields. These are imaged as part of the GOODS NICMOS Survey (GNS). The GOODS field has the multiband coverage needed to estimate good photomentric redshifts, in addition the depth and angular resolution from NICMOS allows us to resolve significantly fainter objects around the primary targets to distances of a few kpc. Thus, for the first time, we can probe the pair fraction history to z $\sim$ 3. We argue that the increase in merger fraction with redshift continues for massive galaxies in our sample out to z $\sim$ 3. The merger fraction shows no sign of levelling off or turning around at $2 < z < 3$, unlike for lower mass systems observed by Conselice et al. (2008). Thus, we find that, complimentary to the star formation results of Bundy et al. (2006) and Pipino et al. (2008), the most massive galaxies appear to form by mergers at a higher rate in the early universe than less massive ones.

Throughout this paper we assume a $\Lambda$CDM Cosmology with: H$_{0}$ = 70 km s$^{-1}$ Mpc$^{-1}$, $\Omega_{m}$ = 0.3, $\Omega_{\Lambda}$ = 0.7, and adopt AB magnitude units.

\section{Data and Observations}

The GOODS NICMOS Survey imaged a total of $\sim$ 8000 galaxies in the F160W (H) band, utilising 180 orbits and 60 pointings of the HST NICMOS-3 camera. These pointings are centred around massive galaxies at z = 1.7 - 3 at 3 orbits depth. Each tile (51.2''x51.2'', 0.203''/pix) was observed in 6 exposures that combine to form images with a pixel scale of 0.1'', and a point spread function (PSF) of $\sim$ 0.3'' full width half maximum (FWHM). See Magee, Bowens \& Illingworth (2007) for details relating to the data reduction procedure. The pointings were chosen and optimised to contain as many high-mass ($M_{*}>10^{11}M_{\odot}$) galaxies as possible. The selection of these is outlined in Conselice et al. (2008, in prep.). These galaxies consist of high redshift galaxies selected by various optical-to-infrared colour techniques (see Papovich et al. 2006, Yan et al. 2004 and Daddi et al. 2007). A total of 82 galaxies were found with $M_{*}>10^{11}M_{\odot}$, with photometric and spectroscopic redshifts between z = 1.7 and z = 3. Limiting magnitudes reached are H = 26.5 (10$\sigma$).

There is a wealth of observational data covering the GOODS field. As such, the masses and photometric redshifts of our sample of massive galaxies are calculated using data from the U-band to the H-band (e.g. Giavalisco et al. 2004). The stellar masses were measured utilising standard multi-colour stellar population fitting techniques, producing uncertainties of $\sim$ 0.2 dex. The details of this procedure can be found in, e.g. Bundy et al. (2006) and Conselice et al. (2007). The stellar masses were calculated assuming a Chabrier initial mass function and by producing model spectral energy distributions (SEDs) constructed by Bruzel \& Charlot (2003) via stellar population synthesis models parametrised by an exponentially declining star formation history. The model SEDs are then fit to the observed SEDs of each galaxy to obtain a stellar mass. Recent work by Conroy, Gunn and White (2008), however, has suggested that current methods for measuring stellar masses may have greater errors than previously thought ($\sim$ 0.3 dex). Although newer methods have been developed utilising different stellar evolution tracks, we find that these newer models do not significantly affect our measured stellar masses, and our stellar masses are accurate to within a factor of a few. For a discussion of this see Trujillo et al. (2007) and Conselice et al. (2008). Due to the steepness of the stellar-mass function at high masses, Poisson errors in measuring the mass of galaxies may lead to more lower mass galaxies being counted in our high mass sample. To model this we ran a Monte-Carlo simulation based on the stellar mass function for high z galaxies in Fontana et al. (2006). We found that $\sim 8\%$ of lower mass galaxies systematically infiltrated our higher mass sample. This effect is relatively small and will not significantly affect our results. Moreover, since we go on to note a dramatic increase in the pair fraction of massive galaxies, when compared to less massive ones, this effect cannot be responsible because contamination of lower mass systems would result in the pair fraction of higher mass galaxies being lower (see e.g. Conselice et al. 2007).

Our photometric redshifts are determined via standard techniques (e.g. Conselice et al. 2008). These contain additional uncertainty on top of the stellar mass issues. Additionally, we find seven spectroscopic redshifts from the literature for our sample. Using the GOODS/VIMOS DR1 (see Popesso et al. 2008) we find three matches with $\frac{\delta z} {1+z} = 0.026$, and four spectroscopic redshifts from a compilation of redshifts from the literature (see Wuyts et al. 2008) giving $\frac{\delta z} {1+z} = 0.034$.

We use SExtractor to make catalogs for all detected galaxies in the GNS, removing stars and spurious detections. We then carefully check by eye that the deblending is accurate and that we are not artificially merging separate galaxies or splitting individual galaxies into parts. For the massive galaxies studied in this paper, we put in particular effort to ensure that the sources are extracted accurately. Further, the choice of studying major mergers with mass ratios of at least 1:4 of the primary source (+/- 1.5 mag.) allowed us to be complete.

\section {Results}
\subsection{Merger Fraction}
\subsubsection{Close Pair Method}

We follow an approach similar to Patton et al. (2000) for calculating major merger fractions for our galaxies based on the pair method. First we separate our galaxies into two redshift bins, with $1.7 < z < 2.3$ for the first group (containing 44 galaxies) and $2.3 < z < 3.0$ for the second group (containing 38 galaxies). The method for assigning potential pairs is relatively straightforward. We assign any galaxy within 30 kpc, in physical units (assuming H$_{0}$ = 70 km s$^{-1}$ Mpc$^{-1}$), of our host galaxy to be a count, if it is within +/- 1.5 of the host galaxy magnitude. To calculate a pair fraction we sum up the number of galaxies within 30 kpc of all of our host galaxies and divide by the total number of galaxies. This would be an accurate pair fraction if we had precise redshifts for all of the galaxies in our sample, and included in the summation only those at the same redshift as the host. As we have redshifts only for our host galaxies, we do not know if any particular galaxy within 30 kpc is a real pair or just foreground or background contamination.

We correct for contamination by calculating the probability of a close galaxy being a pair by chance from our surface number counts. These number counts are taken around the objects to minimise clustering effects. We then subtract this correction from our running pair total. Specifically we calculate:

\begin{equation}
{\rm{corr}} = \int_{m-1.5}^{m+1.5} \rho(m') \times \pi(r_{30kpc}^{2}-r_{5kpc}^{2}) dm'
\end{equation}

\noindent Where $\rho(m')$ is the surface density of galaxies in the survey, in the magnitude range +/- 1.5 of the magnitude of the host galaxy. $r_{30kpc}$ and $r_{5kpc}$ are distances of 30 and 5 kpc from the host galaxy's centre, respectively. Essentially this expression corresponds to calculating how many galaxies in the whole survey are within +/- 1.5 of the magnitude of the host galaxy, and dividing this number by the total area of the survey. We then multiply by the area within the annulus marked out by the radii 5 and 30 kpc, which is done to avoid miscounting due to blending with the host galaxy. This gives an expectation value for the number of galaxies one would expect to see within 30 kpc of the host galaxy purely by chance. Although massive galaxies at high redshifts are highly clustered, our correction is based upon the densities of objects centred around these systems. This value is then subtracted from the number counts to obtain a major pair fraction thus:

\begin{equation}
f_{m} = \frac{1} {N} \sum_{i=1}^{i=N} (\rm{counts_{i} - corr_{i}})
\end{equation}

\noindent Where N is the total number of galaxies in the summation. We use the values of +/- 1.5 for the magnitude range to select major (1:4) mergers only (where we are complete). Furthermore, we adopt the convention of setting 30 kpc as our pair distance to be in line with Patton et al. (2000) and Bundy et al. (2004). This allows us to make fruitful comparisons, and also follows rough theoretical arguments for the likelihood of a major close pair becoming a major merger in a short ($\sim$ 400 Myr) time-scale. The merger fractions derived via close pair methods at d = 30 kpc also have a similar timescale to the merger fractions calculated via CAS methods. For further discussion on this and the characteristic timescales relevant to these two approaches, see Conselice, Yang \& Bluck (2008).

We find a pair fraction of $f_{m}$ = 0.29 +/- 0.06 for the whole sample at $1.7 < z < 3.0$. Within the range $1.7 < z < 2.3$ we calculate a pair fraction $f_{m}$ =  0.19 +/- 0.07, and within the range $2.3 < z < 3.0$ we calculate a pair fraction $f_{m}$ = 0.40 +/- 0.10. These are considerably higher values than those found by Conselice et al. (2008) for similar mass objects at $z < 1.4$. We do not make any correction to transform our pair fractions into merger fractions, unlike Patton et al. (2000) and Rawat et al. (2008). This is because both morphological and close pair methods likely trace merging but have different timescales. A detailed investigation on the differing timescales traced by morphological and pair methods in Conselice, Yang \& Bluck (2008) suggests that the timescale for 30 kpc pairs and morphologically selected mergers is very close. Moreover, we go on to calaculate the pair fractions for the POWIR survey massive galaxies, to directly compare these values to the GNS massive galaxies (see Fig. 1).

N-body simulations from Wetzel et al. (2008) suggest that pair fraction methods may underestimate the number of true major mergers, as pairs at higher separations may also merge. If this is true, it suggests that the real merger fractions may be higher than what we calculate. But, since the calculated merger fractions in this paper are very high, this will not change the thrust of our conclusions and may even make them stronger. Our pair fractions are plotted alongside lower redshift points from the POWIR Survey, with merger fractions estimated via CAS methods (Conselice et al. 2007) and via close pairs (calculated in this paper), and with a local universe value calculated in de Propis et al. (2007) based on morphological methods in Fig. 1. For a detailed explanation of morphological techniques see Conselice et al. (2007).

\subsubsection{Merger Fraction Evolution}

\begin{figure}
\includegraphics[width=0.4\textwidth,height=0.4\textwidth]{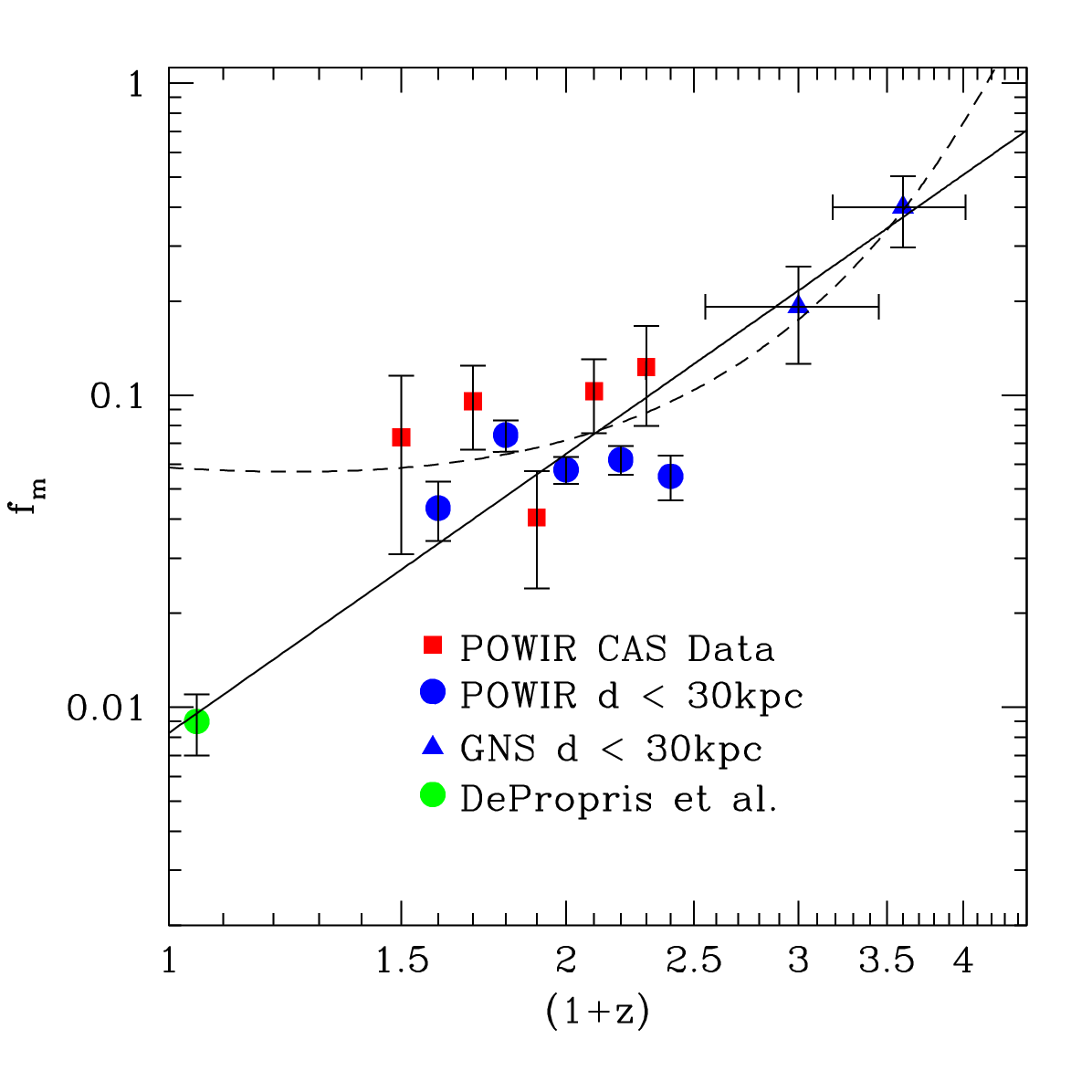}
\caption{The merger fraction evolution of $M_{*}>10^{11}M_{\odot}$ galaxies. The red sqaures are taken from POWIR data, with merger fractions calculated via CAS morphologies (Conselice et al. 2007). The blue circles are taken from POWIR data, with merger fractions calulated via close-pair methods. Blue triangles are taken from the GNS, with merger fractions calculated via close-pair methods. The green circle represents the local universe value (calculated in de Propris et al. 2007). Solid line is a best fit power law, to the high z data, of the form $f_{m}=0.008(1+z)^{3}$, with dotted line being a best fit power law exponential, to the high z data, of the form $f_{m}=0.008(1+z)^{0.3}\exp(1.0(1+z)^{2})$. \label{fig. 1}}
\end{figure}

We plot a fit of the form $f_{m} = f_{0}(1+z)^{m}$ to all POWIR and GNS points (solid line in Fig. 1), and find a best fit to the free parameters of: $f_{0}$ = 0.008 +/- 0.003 and $m$ = 3.0 +/- 0.4. Our value of $f_{0}$ compares very favourably to the accepted local universe value of the merger fraction for massive galaxies of 0.009 +/- 0.002 at z = 0.05 (de Propris et al. 2007). We also fit a power law exponential curve of the form $f_{m}=a(1+z)^{b}\exp(c(1+z)^{2})$ to all POWIR and GNS points (dotted line in Fig. 1) (Carlberg, 1990). We find a best fit to the free parameters of: a = 0.008 +/- 0.002, b = 0.3 +/- 0.2 and c = 1.0 +/- 0.6. There is very poor agreement, however, between the power law exponential prediction for the local universe value and the de Propris et al. (2007) point. This parameterisation also shows no sign of turning over at high z, unlike for the lower mass systems in Conselice et al. (2008). As such, this indicates that the merger fraction continues to increase with redshift out to z $\sim$ 3, implying that higher mass objects are experiencing relatively more mergers than their lower mass counterparts at high redshift, compared to the lower mass data in Conselice et al. (2008). Further, it is also evident that the simple power law curve fits our data better than the more complex power law exponential.

\subsection{Merger Rates}

In order to calculate merger rates from our merger fractions we must assume a time-scale ($\tau_{m}$) over which merging is occuring for galaxies in a major pair. Adopting the simulation results found in Lotz et al. (2008b), we take $\tau_{m}$ = 0.4 +/- 0.2 Gyr for close pairs at d = 30 kpc, and $\tau_{m}$ = 1.0 +/- 0.2 Gyr for morphological CAS measurements. We also need to know the comoving number density, $n(z)$, for $M_{*}>10^{11}M_{\odot}$ galaxies at different redshifts. We calculate this using data from Drory (2005) for the GOODS South field. The typical errors of these values are +/- 20 \% of the value. It is also important to use the galaxy merger fraction ($f_{gm}$) as opposed to the merger fraction calculated above. The galaxy merger fraction relates to the merger fraction ($f_{m}$) by:

\begin{equation}
f_{gm} = \frac{2 \times f_{m}} {1 + f_{m}}
\end{equation}

\noindent This gives the number of galaxies merging as opposed to the number of mergers ($f_{m}$), divided by the number of hosts in the sample. For a more detailed derivation of this and discussion on its application see Conselice (2006). We calculate the merger rate thus:

\begin{equation}
\Re(z) = f_{gm}(z)n(z)\tau_{m}^{-1}
\end{equation}

\noindent where $n(z)$ is the comoving number density of galaxies and $f_{gm}(z)$ is the galaxy merger fraction (calculated above). We calculate this rate at z = 2.6 as $\Re$ $<$ 5 $\times$ 10$^{5}$ Gpc$^{-3}$ Gyr$^{-1}$ and at z = 0.5 as $\Re$ $<$ 1.2 $\times$ 10$^{5}$ Gpc$^{-3}$ Gyr$^{-1}$. These rate calculations are, however, notoriously difficult to perform accurately as there are significant errors associated with the calculation of the comoving number density ($n$), the galaxy merger fraction ($f_{gm}$), and the characteristic timescale for a merging system being detected ($\tau_{m}$). When added in quadrature, the resultant final error on the merger rate is very large. Therefore, unlike for the merger fraction, the major merger rate is consistent with having no redshift evolution. There are two competing variables here: the comoving number density, $n(z)$, which decreases with increasing redshift, and the galaxy merger fraction, $f_{gm}(z)$, which increases with increasing redshift. These appear to combine to roughly negate any redshift evolution of the merger rate $\Re$(z) (Fig. 2).

\begin{figure}
\includegraphics[width=0.4\textwidth,height=0.4\textwidth]{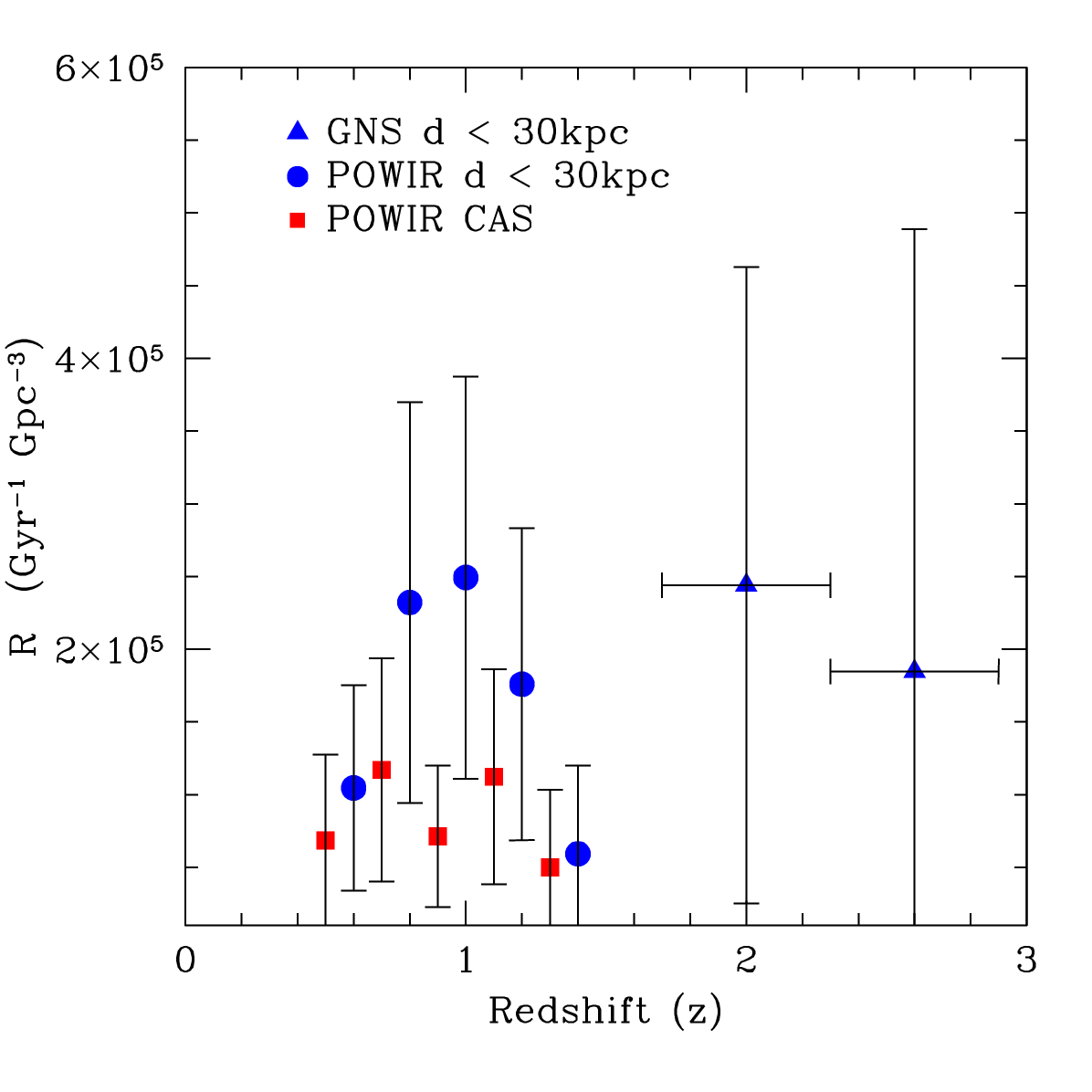}
\caption{The major merger rate, $\Re(z)$. Red squares are taken from POWIR data with merger fractions calculated via morphological CAS methods, blue circles are taken from POWIR data with merger fractions calculated via close pair methods and the blue triangles are high redshift data taken from the GNS with merger fractions derived from close pair methods.}
\end{figure}

\begin{figure}
\includegraphics[width=0.4\textwidth,height=0.4\textwidth]{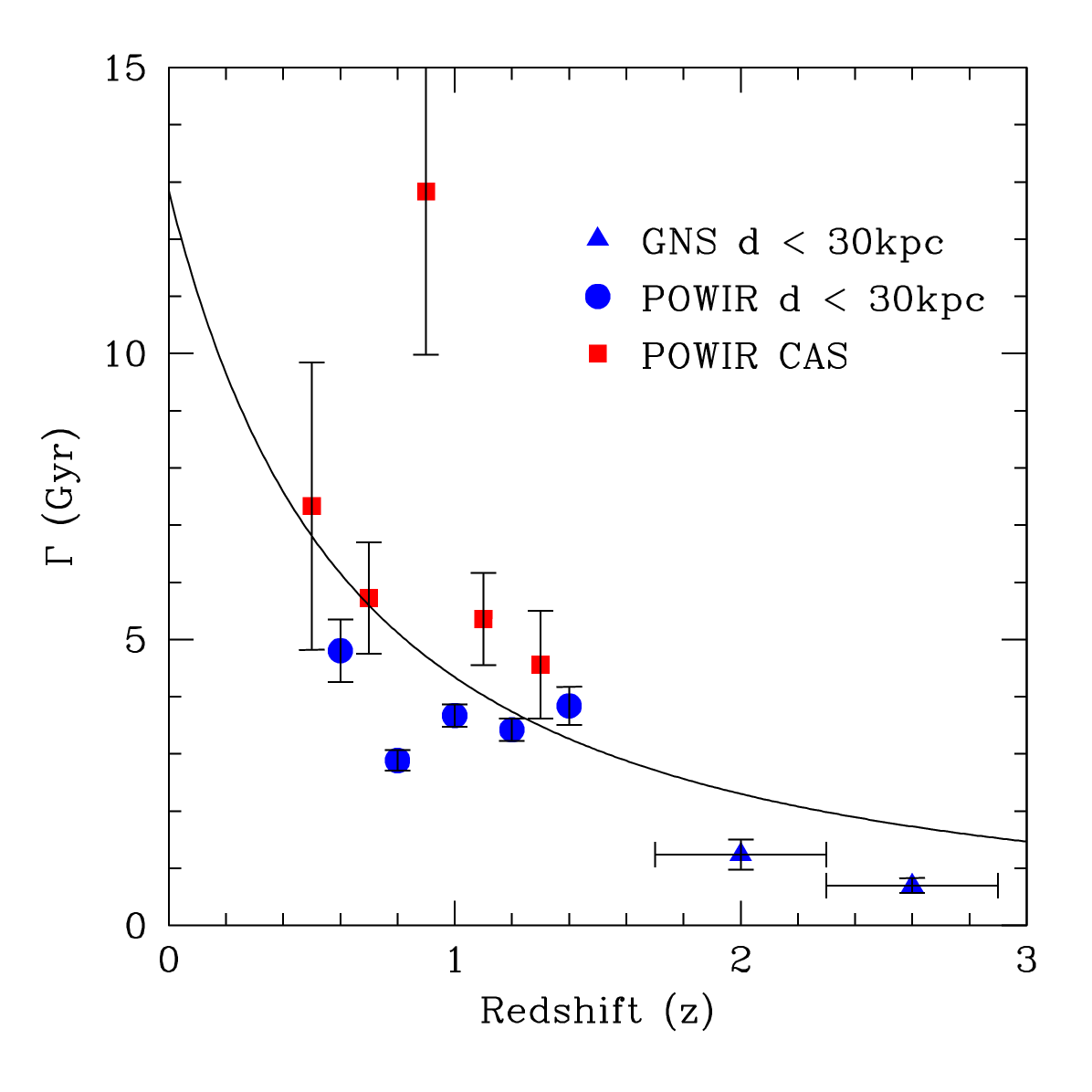}
\caption{The redshift evolution of the average time between mergers $\Gamma = \tau_{m} / f_{gm}$. Red squares are taken from POWIR data with merger fractions calculated via morphological CAS methods, blue circles are taken from POWIR data with merger fractions calculated via close pair methods and the blue triangles are high redshift data taken from the GNS with merger fractions derived from close pair methods. The solid line is the fit $\Gamma = 12(1+z)^{-1.6}$. }
\end{figure}

A useful term may be defined ($\Gamma$) which is essentially the characteristic time between mergers that an average galaxy experiences at a given redshift. It is defind as:

\begin{equation}
\Gamma = \frac{\tau_{m}} {f_{gm}}
\end{equation} 

\noindent This is plotted in Fig. 3. It is clear that $\Gamma$ evolves with redshift from $\sim$ 1.5 Gyr at z = 3 to $\sim$ 12 Gyr at z = 0. Further, we fit a curve of the form $\Gamma = \Gamma_{0}(1+z)^{-p}$, with a best fit to the free parameters of: $\Gamma_{0}$ = 12 +/- 3 Gyr and p = 1.6 + /- 0.4. This represents a steep increase in the time taken for galaxies to merge. We calculate the number of expected mergers ($N_{m}$) for a given galaxy between two redshifts using this power law fit. Specifically we calculate:

\begin{equation}
N_{m} = \int_{t_{1}}^{t_{2}} \frac{1}{\Gamma(z)} dt = \int_{z_{1}}^{z_{2}} \frac{1}{\Gamma(z)} \frac{t_{H}}{(1+z)} \frac {dz}{E(z)}
\end{equation}

\noindent where $\Gamma(z)$ is the characteristic time between mergers, $t_{H}$ is the Hubble time, and the parameter $E(z) = [\Omega_{M}(1+z)^{3}+\Omega_{k}(1+z)^2+\Omega_{\Lambda}]^{-1/2} = H^{-1}(z)$. Calculating this from z = 3 to z = 0, we obtained a value of $N_{m}$ = 1.7 +/- 0.5 major mergers per galaxy, with $\tau_{m}$ = 0.4 Gyr.

\section{Discussion}

Conselice et al. (2008) find that galaxies with masses in the range $8<\log(M_{*}/M_{\odot})<9$ and $9<\log(M_{*}/M_{\odot})<10$ have a peak in their merger fraction at around z = 1.5, with galaxies with stellar masses $\log(M_{*}/M_{\odot})>10$ appearing to peak in their merger fraction later, at around z = 2. In this paper we have found that for galaxies with $\log(M_{*}/M_{\odot})>11$ the merger fraction continues to increase out to z $\sim$ 3, indicating that the peak must occur at z $>$ 3. This provides further evidence that more massive galaxies undergo a greater number of major mergers earlier than less massive systems.

This method of calculating pair fractions is based on noting overdensities of galaxies around host galaxies, based on background counts. However, the clustering of galaxies increases with redshift (see Bell et al. 2006). So, this effect may be partly explained by the correlation function of galaxies increasing with redshift. Since the parameters of this function are poorly known observationally, no attempt has been made to correct for this issue. Consequently, our pair fraction results from both the GNS and POWIR surveys may be interpreted as the 2 point correlation at 30 kpc for these massive galaxies. This clearly evolves and it is likely true that the actual pair fraction evolves in line with this correlation. Contamination from line of sight projection at scales larger than 30 kpc will be minimised by the fact that we calculate our correction around the galaxies being measured. Further, we have recently acquired morphological CAS measurements on the GNS galaxies as well as the POWIR ones. This gives a consistent merger fraction ($f_{m} \sim 0.3$) to the pair methods. We explore this in a forthcoming paper, and elude to the relation between morphologiocal and pair methods in Conselice, Yang \& Bluck (2008).

Despite the large errors associated with calculating merger rates, we have demonstrated that there is little or no evolution in the rate of major mergers with redshift. The maximum evolution permitted by the errors would only amount to increasing the merger rate by a factor of a few. Of greater significance, however, is our calculation of the redshift dependence on the characteristic time between mergers, $\Gamma(z)$. This quantity decreases significantly with redshift. As such, in the early universe (z $\sim$ 3) there was a much shorter time between mergers than there is today.

\section{Summary}

We derive the major merger fractions for massive ($M_{*} > 10^{11} M_{\odot}$) galaxies in the GNS at two redshifts. We find a merger fraction of $f_{m}$ = 0.19 +/- 0.07 at z = 2.0 and $f_{m}$ = 0.40 +/- 0.10 at z = 2.6. This indicates, when compared to data from Conselice et al. (2007), De Propis et al. (2007), and POWIR pair fractions (also calculated in this paper) in Fig. 2, that there is a strong correlation between merger fraction and redshift for the massive galaxies in our sample. Furthermore, we fit a function of the form $f_{m}=f_{0}(1+z)^{m}$ to our data, and the data from Conselice et al. (2007) at lower redshifts for $M_{*} > 10^{11} M_{\odot}$ galaxies, finding a best fit to the two free parameters of: $f_{0}$ = 0.008 +/- 0.003 and $m$ = 3.0 +/- 0.4. This indicates that the merger fraction for massive galaxies continues to increase out to z $\sim$ 3. This implies that more massive galaxies have higher numbers of mergers than less massive ones, since pevious work has found that the merger fraction for lower mass galaxies levels off and declines with increasing redshift in the range z $\sim$ 1.5 - 2.0 (Conselice et al. 2008).

We calculate the major merger rate finding no significant evolution from $\Re$ $<$ 5 $\times$ 10$^{5}$ Gpc$^{-3}$ Gyr$^{-1}$ at z = 2.6 to $\Re$ $<$ 1.2 $\times$ 10$^{5}$ Gpc$^{-3}$ Gyr$^{-1}$ at z = 0.5. Moreover, a steep evolution is ruled out. We also calculate the evolution of the characteristic time between mergers $\Gamma(z)$, finding a rapid decrease from $\sim$ 12 Gyr at z = 0 to $\sim$ 1.5 Gyr at z = 3. This indicates that massive galaxies take longer between individual mergers in the local universe than at high z, but that there are more massive galaxies now than there were in the early universe, so, the rate remains constant. By integrating over the best fit to this curve, we calculate the average number of major mergers a massive ($M_{*} > 10^{11} M_{\odot}$) galaxy would experience between z = 3 and z = 0 is $N_{m}$ = 1.7 +/- 0.5.

We would like to thank our collaborators on the GNS team, particularly Fernando Bruitrago and Ignacio Trujillo, for their work on this project and help in producing this paper, and Samantha Penny for her assistance. We also gratefully acknowledge support from the STFC. \\

REFERENCES \\

\noindent Bell, E. F. et al., 2006, ApJ, 652, 270 \\
Bruzual G. \& Charlot S., 2003, MNRAS, 344, 1000 \\
Bundy K. et al., 2004, ApJL, 601, L123 \\
Bundy K. et al., 2006, ApJ, 651, 120 \\
Carlberg R., 1990, ApJ, 359, L1 \\
Conselice C. J., 2003, ApJS, 147, 1 \\
Conselice C. J., 2006, ApJ, 639, 120 \\
Conselice C. J., et al. 2003, AJ, 126, 1183 \\
Conselice C. J., et al. 2007, MNRAS, 381, 962 \\
Conselice C. J., et al. 2008, MNRAS, 386, 909 \\
Conroy C., Gunn J. E., White M., 2008, submitted to ApJ (arXiv:0809.4261C) \\
Drory N., 2005, ApJ, 619, L131 \\
Daddi E. et al., 2007, ApJ, 670, 156 \\
de Propris et al., 2007, ApJ, 666, 212D \\
Fontana, A. et al., 2006, AA, 5474 \\
Giavalisco M. et al., 2004, ApJ, 600, L93 \\
Lin L. et al., 2004, ApJL, 617, L9 \\
Lin L. et al., 2008, ApJ, 681, 232 \\
Lotz J. M. et al., 2008a, ApJ 672, 177 \\
Lotz J. M. et al., 2008b, MNRAS, in press (arxiv:0805.1246v1) \\
Magee D. K. et al., 2007, Astronomical Data Analysis Software and Systems XVI, 376, 261 \\
Mathis H. et al., 2005, MNRAS, 365, 385 \\
Palpino A. \& Matteucci F., 2008, (arXiv:0805.0793v1) \\
Papovich C. et al., 2006, ApJ, 640, 92 \\
Patton D. R. et al., 2000, ApJ, 536, 153 \\
Popesso P. et al., 2008, arXiv:0802.2930 \\
Rawat A. et al., 2008, ApJ, in press, (arXiv:0804.0078v1) \\
Trujillo I. et al., 2007, MNRAS, 382, 109 \\
Voit M. G., 2004, Rev. Mod. Phys., (arXiv:astro-ph/0410173v1) \\
Wetzel A. R., 2008, ApJ, 683, 1 \\
Wuyts S. et al., 2008, ApJ, in press (arXiv:0804.0615) \\
Yan H. et al., 2004, ApJ, 616, 63 \\

\end{document}